\documentclass{PoS}

\title{Three-Particle Azimuthal Correlations}

\ShortTitle{Three-Particle Azimuthal Correlations}

\author{\speaker{Jason Glyndwr Ulery}\thanks{}\\
        Purdue University\\
        E-mail: \email{ulery@physics.purdue.edu}}

\abstract{
Two-particle azimuthal correlations in central Au+Au collisional at RHIC have revealed a broadened away-side structure, with respect to perpherial Au+Au, {\it pp}, and d+Au.  This could be explained by different physics mechanisms such as: large angle gluon radiation, deflected jets, $\hat{C}$erenkov gluon radiation, and conical flow generated by hydrodynamic shock-waves.  We can discriminate the scenarios with conical emission, $\hat{C}$erenkov radiation and conical flow, from the other mechanisms though three-particle correlations.  In addition, the associated particle $p_T$ dependence can be used to distinguish conical flow from simple $\hat{C}$erenkov gluon radiation.  We will discuss three-particle correlation analyses that have been performed at RHIC and what can be done at the LHC.
}

\FullConference{High-pT physics at LHC\\
		 March 23-27 2007\\
		 University of Jyv\"askyl\"a, Jyv\"askyl\"a, Finland}

\begin{document}

\section{Introduction}
Heavy-ion collisions create a medium that could be the quark gluon plasma (QGP).  Jets and jet-correlations can be used to study this medium through the effects of the medium on the jet and the effects of the jet on the medium.  Jets make a good probe because their properties can be calculated in the vacuum with perturbative quantum chromodynamic (pQCD).  Two-particle jet-like azimuthal correlations have shown a broadened (with respect to what is seen in {\it pp} and d+Au and perpherial Au+Au collisions) or even double-humped\cite{star2p,phenix2p} away-side structure in Au+Au central collisions (see Fig.~\ref{fig:me1}).  The broadening of the away-side structure is consistent with different physics mechanisms including:  large angle gluon radiation \cite{gluon1,gluon2}, jets deflected by radial flow\cite{deflected} or preferential selection of particles due to path-length dependent energy loss\cite{path}, hydrodynamic conical flow generated by Mach-cone shock waves \cite{mach1,mach2}, and \v{C}erenkov radiation \cite{cerenkov1,cerenkov2}.  Three-particle correlations can be used to differentiate the mechanisms with conical emission, Mach-cone and \v{C}erenkov gluon radiation, from other physics mechanisms.  Three different 3-particle correlation analyses have been done at RHIC.  Two of these analyses are azimuthal 3-particle correlations and will be discussed here.  The third analysis makes use of the full 3D information available and is discussed in detail else where in these proceedings\cite{phenix3p} and elsewhere\cite{phenix3p2}.

\section{Three-Particle Cumulant}

\begin{figure}[htbp]
	\centering
		\includegraphics[width=.50\textwidth]{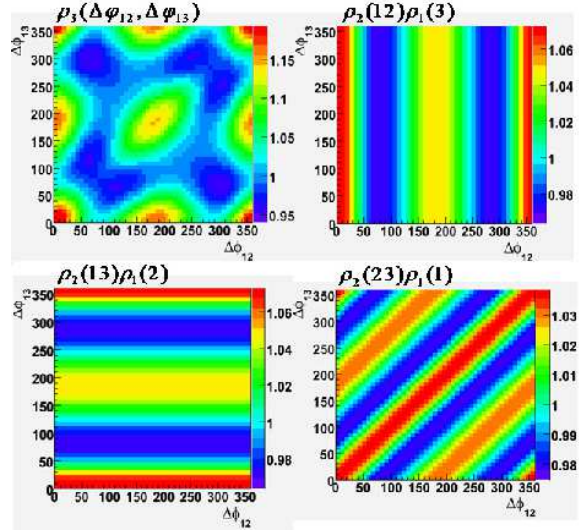}
			\vspace*{-0.35cm}
	\caption{Cumulant raw signal and backgrounds which contain structure.  Top Left: Cumulant raw signal, $\rho_{3}(\Delta\phi_{1},\Delta\phi_{2})$.  Top right: $\rho_{2}(\Delta\phi_{1})\rho_{1}(\phi_{2})$.  Bottom Left: $\rho_{2}(\Delta\phi_{2})\rho_{1}(\phi_{1})$.  Bottom Right:  $\rho_{2}(\phi_{1}-\phi_{2})\rho_{1}(\phi_{T})$.  All results are preliminary.}
	\label{fig:cumulant}
\end{figure}

The 3-particle cumulant has previously shown preliminary results\cite{claudeQM} and been rigorously described in\cite{claudeMethod}.  This method has the advantage that it is independent of any model.  This method is performed by computing 1-,2-, and 3-particle densities and then computing the quantity:
\begin{eqnarray}
C_{3}(\Delta\phi_{1},\Delta\phi_{2})&=&\rho_{3}(\Delta\phi_{1},\Delta\phi_{2})-\rho_{2}(\Delta\phi_{1})\rho_{1}(\phi_{2})-\rho_{2}(\Delta\phi_{2})\rho_{1}(\phi_{1})\nonumber \\
&&-\rho_{2}(\phi_{1}-\phi_{2})\rho_{1}(\phi_{T})-\rho1(\phi_{T})\rho1(\phi_{1})\rho1(\phi_{2})
\label{cumequation}
\end{eqnarray}
where $\rho_{1}$,$\rho_{2}$,and $\rho_{3}$ are the 1-, 2-, and 3-particle densities respectively, $\phi_{T}$, $\phi_{1}$, and $\phi_{2}$ are the azimuthal angles of the trigger and 2 associated particles respectively, and $\Delta\phi_{1}=\phi_{T}-\phi_{1}$ and $\Delta\phi_{2}=\phi_{2}-\phi_{1}$.  Figure~\ref{fig:cumulant} shows the terms $\rho_{3}(\Delta\phi_{1},\Delta\phi_{2})$, $\rho_{2}(\Delta\phi_{1})\rho_{1}(\phi_{2})$, $\rho_{2}(\Delta\phi_{2})\rho_{1}(\phi_{1})$, and $\rho_{2}(\phi_{1}-\phi_{2})\rho_{1}(\phi_{T})$.

\begin{figure}[htbp]
	\centering
		\includegraphics[width=1.00\textwidth]{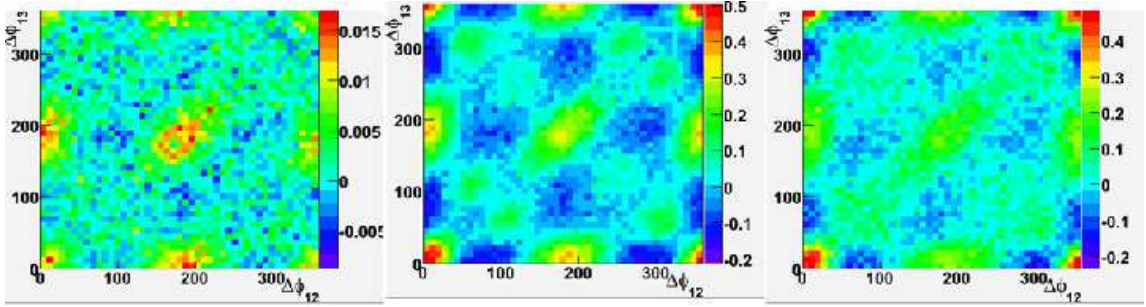}
			\vspace*{-0.35cm}
	\caption{Three-particle cumulants for Au+Au collisions in centrality bins 50-80\% (left), 10-30\% (middle), and 0-10\% (right) with $3<p_{T}^{Trig}<4$ and $1<p_{T}^{Assoc}<2$ GeV/c.  All results are preliminary.}
	\label{fig:cumulant2}
\end{figure}

If the events are Possion the 2-particle correlations are fully removed from the 3-particle cumulant and only the correlations of 3 or more particles remain.  Therefore the presence of a signal signifies the existence of correlations of 3 or more particles.  Figure~\ref{fig:cumulant2} shows the 3-particle cumulant for three different centrality bins in Au+Au collisions.  The cumulants are for trigger particles of $3<p_{T}<4$ GeV/c with two associated particles of $1<p_{T}<2$ GeV/c.  All particles are charged particles within the STAR TPC.  A signal is seen for all centrality bins signifying the presence of correlations of three or more particles, under the assumption the events are Possion.  A centrality dependence is seen for the 3-particle cumulant.  The away-side structure are inconsistent with those predicted for global momentum conservation in the cumulant\cite{momcon}.  Any further interperation requires invoking a model and studying the effects of components of the model on the cumulant.  To extract additional information we will proceed to discuss a model dependent analysis.

\section{Model Dependent Analysis}

This analysis strives to extract the 3-particle jet-like correlations.  The method has been described in detail\cite{myMethod} and previously reported preliminary results\cite{myQM, myHP, claudeQM}.  This method makes a few assumptions.  The first is the event can be composed into two components, particles that are jet-like correlated with the trigger particle and particles that are not.  The other assumptions are that the background in the 2-particle correlation is ZYA1 (zero yield at 1) and that the number of associated pairs in the three-particle signal is the square of the number of associated particles in the 2-particle signal.  Results are shown for charged trigger particles of $3<p_{T}<4$ GeV/c correlated with charged associated particles for $1<p_{T}<2$ GeV/c taken in the STAR TPC.  Results are for {\it pp}, d+Au, and Au+Au collisions at $\sqrt{s_{NN}}$=200 GeV.

\begin{figure}[htbp]
	\centering
		\includegraphics[width=1.0\textwidth]{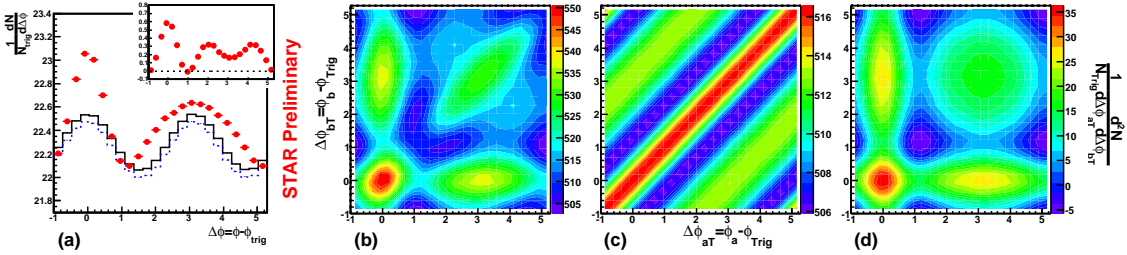}
			\vspace*{-0.4cm}
	\caption{(a) Raw 2-particle correlation (points), background from mixed events with flow modulation added-in (solid) and scaled by ZYA1 (dashed), and background subtracted 2-particle correlation (insert).  (b) Raw 3-particle correlation, (c) soft-soft background, $\beta \alpha^{2}B_{3}^{inc}$ and (d) hard-soft background + trigger flow, $\hat{J}_{2}\otimes \alpha B_{2}^{inc}$ + $\beta \alpha^{2}B_{3}^{inc,TF}$.  See text for detail.  Plots are from ZDC-triggered 0-12\% Au+Au collisions at $\sqrt{s_{NN}}$=200 GeV/c.}
	\label{fig:me1}
\end{figure}

Figure~\ref{fig:me1}b shows the raw 3-particle correlation signal in $\Delta\phi_{1}=\phi_{1}-\phi_{T}$ and $\Delta\phi_{2}=\phi_{2}-\phi_{T}$.  Combinatorial backgrounds must be removed to obtain the genuine 3-particle jet-like correlation signal.  One source of background, the hard-soft background, is obtained by folding the background subtracted 2-particle correlation signal (Fig.~\ref{fig:me1}a, minipanel), $\hat{J_{2}}$, with the 2-particle background, $a B_{2}^{inc}$, (Fig.~\ref{fig:me1}a, dashed line).  This background is used to removed instances of only one particle having a jet-like correlation with the trigger particle.  The 2-particle background is constructed from event mixing with the flow modulation added pairwise from the average of the measurements based on the reaction plane and 4-particle cumulant methods\cite{star2p}.  For the $v_{4}$ contribution the parameterization $v_{4}=1.15 v_{2}^{2}$ is used and was obtained from data\cite{flow}.  The 2-particle background is normalized by factor $a$ such that the background subtracted signal is ZYA1.  To take care of the instances where neither associated particle is correlated with the trigger but are correlated with each other a trigger particle from one event is mixed with associated particles from another inclusive event, events in the same centrality widow.  This is referred to as the soft-soft term, $B_{3}^{inc}$.  Since the two associated particles are from the same event, this background contains all of the correlations between the two associated particle that are independent of the trigger particle, including minijets, other jets in the events, decays, and flow.

There is additional flow that is not accounted for by the soft-soft term.  The associated particles are also correlated with the trigger via flow.  The trigger flow is added in triplet-wise from mixed events where the trigger and associated particles are all from different inclusive events.  The $v_{2}$ and $v_{4}$ values are obtained the same way as in the 2-particle background.  The total number of triplets is determined from inclusive events.  We shall refer to the background from trigger flow as $B_{3}^{inc,TF}$.  The total background is then, $\hat{J}_{2}\otimes \alpha B_{2}^{inc}$ + $\beta \alpha^{2}(B_{3}^{inc}+B_{3}^{inc,TF})$ where $a^{2}$ accounts for the multiplicity bias from requiring a trigger particle and $b$ accounts for the effect of non-Poisson multiplicity distributions.  The normalization factor $b$ is obtained such that the number of associated pairs in the background subtracted jet-like 3-particle correlation signal equals the square of the number of associated particles in the background subtracted jet-like 2-particle correlation signal.  

\begin{figure}[htbp]
	\centering
		\includegraphics[width=.95\textwidth]{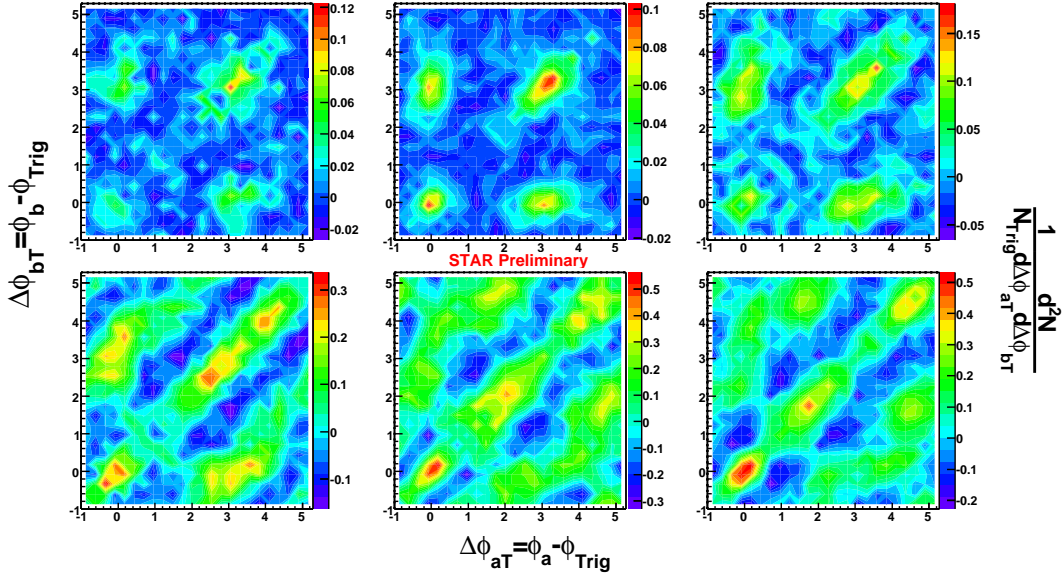}
			\vspace*{-0.3cm}
	\caption{Background subtracted 3-particle correlations for {\it pp} (top left), d+Au (top center), and Au+Au 50-80{\%} (top right), 30-50{\%} (bottom left), 10-30{\%} (bottom center), and ZDC triggered 0-12{\%} (bottom right) collisions at $\sqrt{s_{NN}}$=200 GeV/c.}
	\label{fig:me2}
\end{figure}

Figure~\ref{fig:me2} shows the background subtracted results for jet-like 3-particle correlations.  The $pp$ and d+Au results are similar and show on-diagonal broadening qualatively consistent with $k_{T}$ broadening.  Au+Au collisions show additional on-diagonal broadening. In the more central Au+Au collisions, there is an off-diagonal structure at about $\pi\pm$1.45 radians. This structure is consistent with conical emission and increases in magnitude with centrality.

\begin{figure}[htbp]
	\centering
		\includegraphics[width=1.0\textwidth]{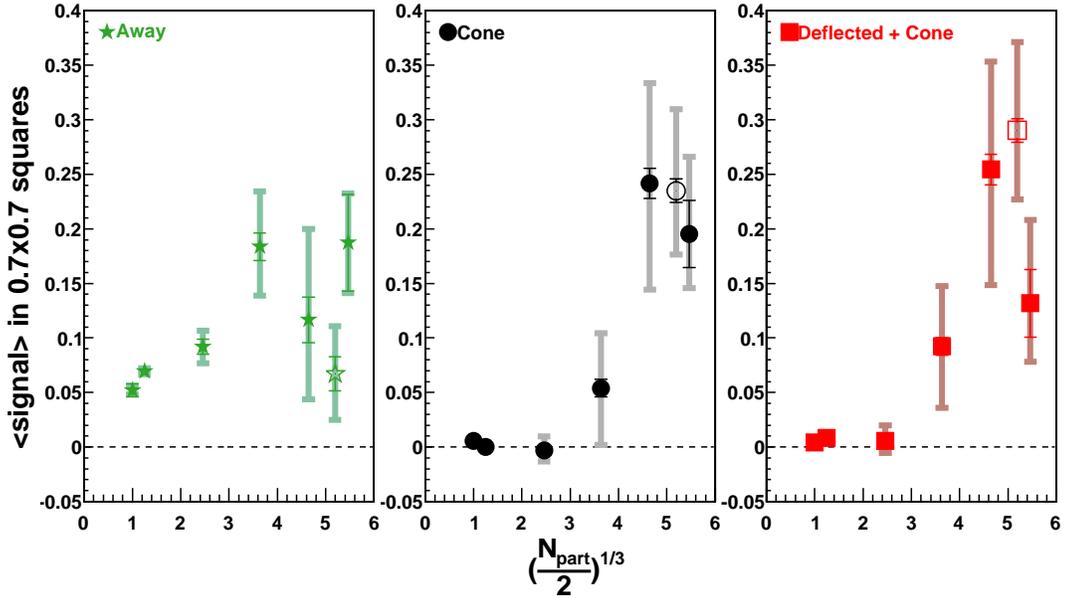}
			\vspace*{-0.0cm}
	\caption{Average signals in 0.7 $\times$ 0.7 boxes at ($\pi$,$\pi$), left, ($\pi\pm1.45$,$\pi\mp1.45$), center, and ($\pi\pm1.45$,$\pi\pm1.45$), right.  Solid error bars are statistical and shaded are systematic.  $N_{part}$ is the number of participants.  The ZDC 0-12\% points (open symbols) are shifted to the left for clarity.}
	\label{fig:me3}
\end{figure}

\begin{figure}[htbp]
	\centering
		\includegraphics[width=1\textwidth]{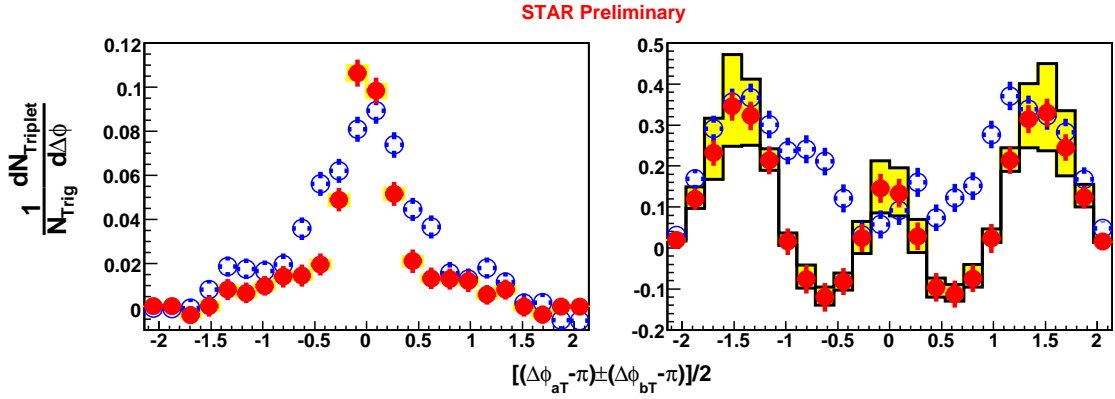}
			\vspace*{-0.0cm}
	\caption{Away-side projections of a strip of width 0.7 radians for (left) d+Au and (right) 0-12\% ZDC Triggered Au+Au.  Off-diagonal projection (solid) is $(\Delta\phi_{aT}-\Delta\phi_{bT})/2$ and on-diagonal projection (open) is $(\Delta\phi_{aT}+\Delta\phi_{bT})/2-\pi$. Shaded bands are systematic errors on the on-diagonal projections.  The systematic errors on the off-diagonal projections are not shown for clarity.}
	\label{fig:me4}
\end{figure}

Figure~\ref{fig:me3} shows the centrality dependence of the average signal strengths in different regions.  The off-diagonal signals increase with centrality and significantly deviate from zero in central Au+Au collisions.  The systemaitc errors are semi-correlated between centralities and data sets.  In figure~\ref{fig:me4} the off-diagonal projection is shown in the solid symbols and the on-diagonal projection is shown in open symbols.  There is additional contribution along the diagonal relative to the off-diagonal.  The locations of the off-diagonal signals were determined from a fit to a central Gaussian and symmetric side Gaussians to a strip projected to the off-diagonal (Fig.~\ref{fig:me4}) and were found to be about 1.45 radians from $\pi$.  This could be due to any combination of jets deflected by flow, path-length dependent energy loss, large angle gluon radiation, and conical emission from Mach-cones\cite{renk}.  Differentiation of the contribution to the on-diagonal projection from these processes cannot be determined experimentally and would require information from theoretical models.

\section{Summary}
Three different 3-particle analyses are being perused at RHIC, two of which are reported on in these proceedings.  The RHIC results show the existence of 3 or more particle correlations in the cumulant analysis with only the assumption of Poisson statistics.  Additionally model dependent results show a signal consistent with conical emission in central Au+Au collisions at about $\pi\pm1.45$ radians.  All three of these analyses can be performed at the LHC.  Additionally 4 or more particle correlations may be possible and may give additional information.  Also at the ALICE detector 3-particle correlations with identified particles should be possible.

\end{document}